\documentclass[draftthesis,tocnosub,noragright,centerchapter,fullpagesingle,12pt]{uiuc_csthesis21}

\makeatletter

\usepackage{setspace}  
\usepackage[numbers, sort]{natbib}  
\usepackage{url}  
\usepackage{tikz}
\usetikzlibrary{shapes.geometric, arrows.meta, positioning}
\usepackage[a4paper,margin=1in]{geometry} 
\usepackage{array}
\usepackage{tabularx} 
\usepackage{booktabs}
\usepackage{lscape}  
\def\fillandplacepagenumber{
	\par
	\pagestyle{empty}
	\vbox to 0pt{\vss}
	\vfill
	\vbox to 0pt{
		\baselineskip 0pt
		\hbox to \linewidth{\hss}
		\baselineskip\footskip
		\hbox to \linewidth{\hfil\thepage\hfil}\vss
	}
}

\usepackage{graphicx}  
\usepackage{caption}
\usepackage{booktabs}  
\usepackage{amsfonts}
\usepackage{amsmath}
\usepackage{amssymb}
\usepackage{amstext}
\usepackage{amsthm}

\theoremstyle{definition}

\usepackage{listings}  
\usepackage[ruled]{algorithm2e}  
\numberwithin{algocf}{chapter}     

\msthesis

\title{LLM2IR: Simple Unsupervised Contrastive Learning Makes Long-context LLM Great Retriever}
\author{Xiaocong Yang}
\department{Computer Science}

\advisor{Professor Chengxiang Zhai}

\begin{document}

%

%
\maketitle

\parindent 1em%

\frontmatter

%
\newcommand{\model}{\textsc{LLM2IR}}

\begin{abstract}
Modern dense information retrieval (IR) models usually rely on costly large-scale pretraining. In this paper, we introduce \model, an efficient unsupervised contrastive learning framework to convert any decoder-only large language model (LLM) to an information retrieval model. Despite its simplicity, the effectiveness is proven among different LLMs on multiple IR benchmarks including LoCo, LongEmbed and BEIR. We also find that models with a longer context length tend to have a stronger IR capacity by comparing task performances of models in the same model family. Our work not only provides an effective way to build IR models on the state-of-the-art LLMs, but also shed light on the relationship between information retrieval ability and model context length, which helps the design of better information retrievers.
\end{abstract}

%

%
\begin{acknowledgments}
I want to express my sincere gratitude to my advisor, Professor Chengxiang Zhai, for his kind guidance on my research, as well as Professor Wenhu Chen from University of Waterloo for the help on this thesis project.
\end{acknowledgments}

%
\tableofcontents

\mainmatter

%
\chapter{Introduction}
Information Retrieval (IR) lies at the heart of the modern information society, powering everything from open-domain question-answering assistants and academic search engines to e-commerce product ranking and large-scale retrieval-augmented generation (RAG) systems. At its core, IR attempts to estimate the relevance of a document $d$ to a user information need expressed by a query $q$, returning an ordered list that maximizes the user’s utility under strict latency and memory constraints.

Over the past five decades, the dominant paradigm has evolved through three major eras:

	1.	Term-matching era (1970 – 2010). Vector-space models, TF–IDF weighting and probabilistic frameworks such as BM25 enabled efficient inverted-index search at web scale. Despite their robustness and interpretability, these sparse models depend on lexical overlap and struggle with synonymy, paraphrase and minor spelling variation—collectively known as vocabulary mismatch.
    
	2.	Representation-learning era (2013 – 2020). Inspired by word2vec and BERT \cite{devlin2019bertpretrainingdeepbidirectional}, dense dual-encoder retrievers learned to embed queries and documents into a shared semantic space, narrowing the lexical gap and improving recall. However, dual encoders usually require millions of labeled query–document pairs — curated for web search, community QA or product recommendations — to achieve competitive performance. The annotation cost and domain drift hinder adoption in specialized corpora such as biomedical literature or legal opinions.
    
	3.	LLM-centric era (2020 – present). The arrival of decoder-only Large Language Models (LLMs) such as GPT-4 \cite{openai2024gpt4technicalreport}, trained at trillion-token scale, has blurred the boundary between retrieval, reasoning and generation. These models encode far richer linguistic knowledge than task-specific dual encoders, raising a natural question: Can a single LLM backbone simultaneously retrieve and generate, eliminating the need to maintain two disjoint model families?

\section{Problem Statement}

This thesis starts from a minimalist premise:

	With nothing more than raw unlabeled text and lightweight parameter adaptation, one can repurpose an off-the-shelf decoder-only LLM into a competitive dense retriever that scales to very long documents.

Testing this hypothesis involves three intertwined sub-questions:

	1.	Can unsupervised contrastive learning alone suffice?
    
	2.	What simple data augmentations best approximate real query–document pairs?
    
	3.	Do the resulting models generalise beyond the training corpus to heterogeneous benchmarks such as LoCo, LongEmbed and BEIR?

\section{Proposed Approach: LLM2IR}

We introduce LLM2IR, a fully unsupervised framework that converts any decoder-only LLM into an IR model using only two ingredients:

	•	Random cropping for positives. Two slices from the same document act as semantically equivalent views.
    
	•	BM25-mined hard negatives. For each anchor slice, we retrieve K superficially similar — but semantically distinct — documents, ensuring training focuses on fine-grained relevance rather than topical keywords alone.

\section{Empirical Findings}

Extensive experiments reveal three striking patterns:

	•	Simplicity beats complexity. LLM2IR exceeds LLM2Vec, despite skipping bidirectional attention re-wiring—and rivals supervised E5-Mistral on both long- and short-context benchmarks (Tables 4.4–4.7).
    
	•	Context headroom matters. For two model pairs differing only in window size (Phi-3 4 k vs 128 k; Mistral-8 k vs 128 k-YaRN), the longer variant yields an average 6 - 8 point gain in nDCG@10—even when all inputs are truncated to the shorter window (Chapter 5).
    
	•	Attention-edge cliff. Re-plotting LongRope’s Passkey results shows an about 40\% recall drop within the last 10\% of the window. We hypothesise under-training of high-index rotary dimensions as a root cause (Section 6.2).

\section{Contributions}

This thesis therefore contributes:

	1.	A reproducible, hardware-friendly recipe for unsupervised LLM retrieval conversion.
    
	2.	The first systematic study of attention-window effects on IR accuracy across multiple model families.
    
	3.	A practical guideline: choose a retriever whose maximum context length is at least 2 × the longest expected document chunk to avoid sudden degradation.

\section{Thesis Roadmap}

	•	Chapter 2 revisits sparse and dense IR foundations, emphasising efficiency-effectiveness trade-offs.
    
	•	Chapter 3 surveys LLM architectures, emergent abilities and their intersection with IR.
    
	•	Chapter 4 details the LLM2IR methodology, implementation and main experimental results.
    
	•	Chapter 5 quantifies the influence of context length and investigates the attention-edge cliff.
    
	•	Chapter 6 outlines future directions in data augmentation and theoretical analysis of positional under-training.
    
	•	Chapter 7 concludes with key takeaways and prospective research avenues.

By the end of this thesis, we show that minimal unsupervised finetuning unlocks the retrieval potential of large decoder-only models, challenging the notion that dense IR and LLM reasoning must remain separate silos. In doing so, we hope to inspire a new generation of unified, lightweight and long-context retrieval solutions for both academia and industry.
\chapter{Background of Information Retrieval(IR)}
\section{Overview}

Information Retrieval (IR) is the field concerned with locating and ranking relevant information from large collections of unstructured or semi-structured data, typically in response to a user’s query. Unlike traditional database systems that focus on exact matches to structured queries, IR emphasizes finding documents that best satisfy an information need, even when the match is partial, approximate, or based on broader notions of relevance. A classic example is web search engines like Google, where a few keywords must quickly retrieve the most pertinent web pages from billions of documents.

\tikzstyle{data}  = [rectangle, minimum width=4cm, minimum height=1cm, text centered, draw=black, fill=blue!10]
\tikzstyle{arrow} = [thick,->,>=stealth]

\begin{figure}[htbp]
  \centering
  \begin{tikzpicture}[node distance=1.8cm]
    \node (query)    [data, fill=orange!20]                   {User Query};
    \node (docs)     [data, below=of query]                   {Top-K Docs};
    \node (reranked) [data, below=of docs]                    {Re-ranked Docs};
    \node (answer)   [data, below=of reranked, fill=green!20] {Final Answer};

    \node (col) [data, left=of docs, fill=gray!20]            {Documents};

    \draw [arrow] (query)    -- node[right]{\textbf{Retriever}} (docs);
    \draw [arrow] (col)      --                                          (docs);
    \draw [arrow] (docs)     -- node[right]{\textbf{Reranker}}  (reranked);
    \draw [arrow] (reranked) -- node[right]{\textbf{Reader}}    (answer);
  \end{tikzpicture}
  \caption{Information Retrieval pipeline: the query is matched against a document collection by the Retriever to produce Top-K Docs, which are then reordered by the Reranker and finally consumed by the Reader to yield the Final Answer.}
  \label{fig:ir_pipeline}
\end{figure}
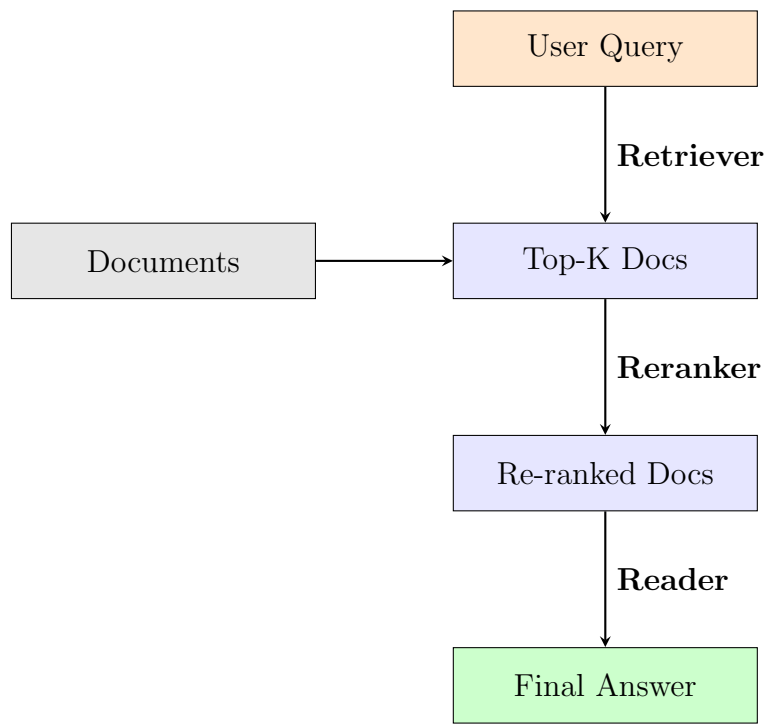

Information Retrieval (IR) systems today are often structured in multiple stages to balance efficiency and accuracy. The first stage is the \textbf{retriever}, which quickly scans a massive document collection to identify a smaller set of candidate documents likely to be relevant to the user’s query. The retriever stage reduces the search space dramatically — from millions or billions of documents down to a manageable few hundred or thousand. Following retrieval, a \textbf{reranker} examines the candidate documents more carefully to produce a finer-grained ranking. While the retriever works fast with relatively simple representations, the reranker applies more sophisticated and computationally intensive models, such as cross-encoders (which jointly encode the query and document) or large transformer-based architectures (e.g., BERT, RoBERTa). In some systems, particularly open-domain question answering (QA) or complex retrieval pipelines, a third stage called the \textbf{reader} is introduced. Rather than simply ranking documents, the reader deeply processes the top-ranked documents to extract specific answers, generate summaries, or perform reasoning. 

Evaluating an IR system involves measuring how effectively it retrieves relevant information, typically using metrics like precision, recall, and Normalized Discounted Cumulative Gain (NDCG). As the volume and complexity of information continue to grow, IR has evolved to integrate advances in machine learning, natural language processing, and user modeling, pushing the frontier toward ever more intelligent and personalized retrieval experiences.

\section{Sparse IR Model}
Sparse IR models are a class of retrieval systems that represent documents and queries using high-dimensional, sparse vectors — where most entries are zero. These models rely on the principle that the presence (or weighted frequency) of specific terms in documents and queries carries semantic significance. The most classic and widely used sparse IR model is BM25 \cite{10.1561/1500000019}, which scores documents based on exact keyword overlaps and their term-level importance in the corpus:

\begin{equation}
    BM25(q, d) = \sum_{t \in q} {IDF}(t) \cdot \frac{f(t, d) \cdot (k_1 + 1)}{f(t, d) + k_1 \cdot \left(1 - b + b \cdot \frac{|d|}{avgdl}\right)}
\end{equation}

\begin{equation}
    IDF(t) = \log \left( \frac{N - n_t + 0.5}{n_t + 0.5} + 1 \right)
\end{equation}

where $q$ is the user query, $d$ is the document, $f(t, d)$ is the frequency of term $t$ in document $d$, $|d|$ is the length of document $d$, $avgdl$ is the average document length in the collection, $k_1$ and $b$ are hyperparameters; $N$ is the number of documents in the corpus, and $n_t$ is the number of documents containing term $t$. Sparse models build inverted indexes that map terms to the documents in which they appear, enabling extremely fast lookups and scalable retrieval across billions of documents. Despite their simplicity, sparse models remain dominant in many real-world search engines due to their efficiency, interpretability, and robustness in low-resource or multilingual settings. However, they struggle with lexical mismatch — failing to retrieve relevant documents that use synonyms or paraphrases.

To address this, modern work explores ways to enhance sparse models through learned sparse representations, such as SPLADE \cite{DBLP:journals/corr/abs-2107-05720}, DeepCT \cite{dai2019contextawaresentencepassagetermimportance} and SparTerm \cite{bai2020spartermlearningtermbasedsparse}, which predict sparse document term weights using BERT-style models. These models aim to combine the best of both worlds: retaining efficient inverted-index-based search while capturing deeper semantic associations. Sparse IR remains a crucial baseline and component in hybrid systems that combine sparse and dense retrieval.

\begin{figure}[htbp]
  \centering
  \includegraphics[width=0.99\linewidth]{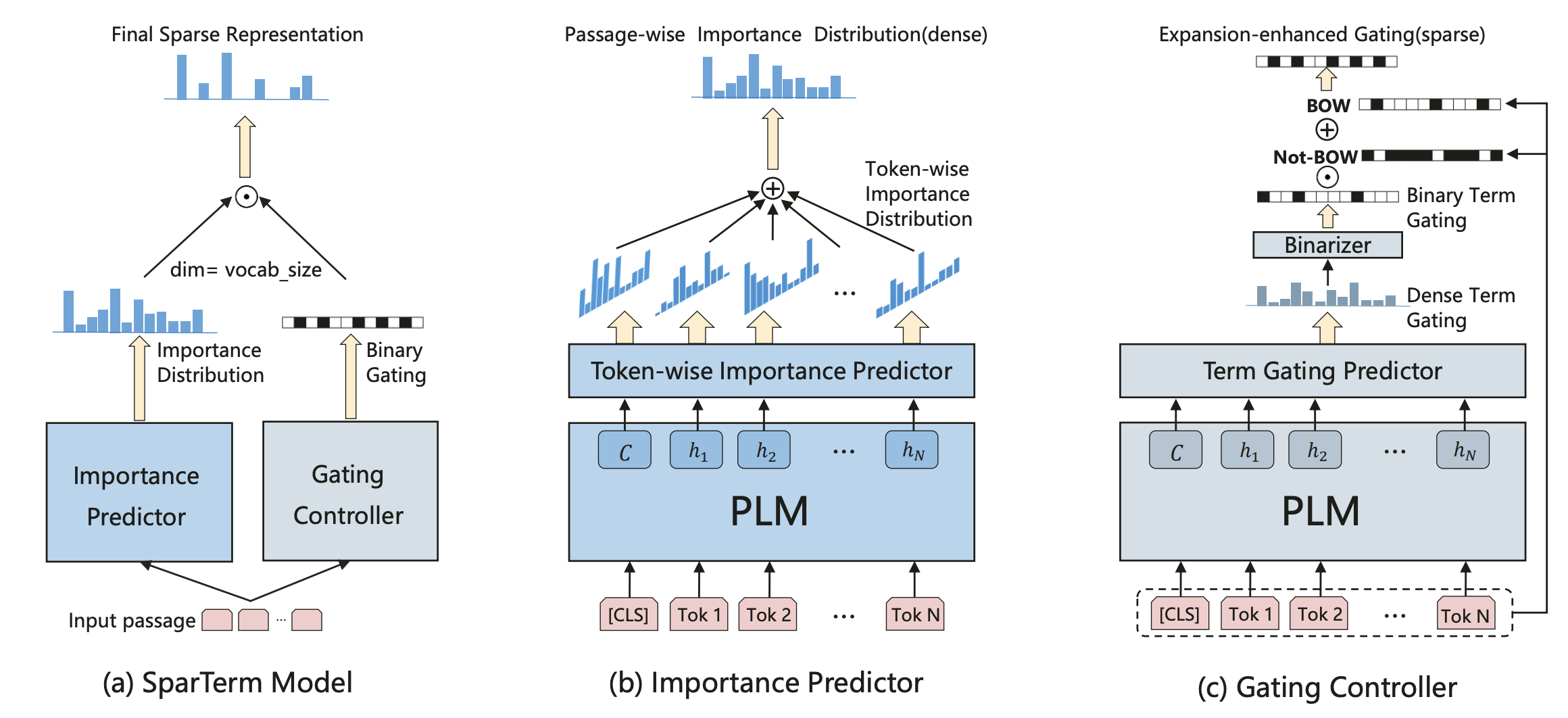}
  \caption{Main framework of SparTerm \cite{bai2020spartermlearningtermbasedsparse}. Pretrained LMs are used as importance predictor and term gating predictor to produce sparse vector representations. }
  \label{fig:sparterm}
\end{figure}

\section{Dense IR Model}
Dense Information Retrieval represents a shift from traditional keyword-based retrieval methods toward semantically richer models that encode both queries and documents as dense vectors in a shared continuous space. These models are typically built upon pretrained language models such as BERT \cite{devlin2019bertpretrainingdeepbidirectional}, RoBERTa \cite{liu2019robertarobustlyoptimizedbert}, or newer transformer-based architectures, which can capture deeper linguistic and contextual relationships than sparse term-frequency methods. The motivation behind dense IR stems from the limitations of sparse models like BM25, which perform exact term matching and often fail when there’s a lexical gap — i.e., when the query and relevant document use different wording (e.g., “car” vs. “automobile”).

The core architecture of a dense retrieval system is typically a dual encoder or bi-encoder model. In this framework, the query and document are each passed through separate (but often parameter-shared) neural encoders to produce fixed-size embeddings. Relevance between a query and a document is then computed using a similarity function, usually dot product or cosine similarity. Because this architecture allows for independent encoding, it enables fast retrieval over large corpora using Approximate Nearest Neighbor (ANN) search algorithms such as FAISS, HNSW, or ScaNN. This decoupling makes dense IR particularly attractive for real-time applications where latency is critical.

Dense IR models are trained to align semantically similar query-document pairs through contrastive learning objectives. One of the most widely used training paradigms is in-batch negative sampling, where positive query-document pairs are contrasted against other documents in the same mini-batch. More advanced training techniques include hard negative mining (selecting especially confusing negatives to improve robustness), multi-vector representation (e.g., ColBERT \cite{khattab2020colbertefficienteffectivepassage}), and knowledge distillation from larger cross-encoders. Pretrained dual encoders such as DPR \cite{karpukhin2020densepassageretrievalopendomain} and E5 \cite{wang2024textembeddingsweaklysupervisedcontrastive} have demonstrated strong performance across a range of retrieval benchmarks including MSMARCO, Natural Questions, and BEIR.

Dense IR models offer significant benefits over sparse models in several key areas. They are inherently better at semantic matching, making them suitable for tasks where vocabulary varies (e.g. question answering, paraphrase retrieval). They also perform well in multilingual and low-resource settings, where traditional inverted indexes break down. Dense IR plays a critical role in open-domain QA systems, where it’s used to retrieve candidate passages from which answers are extracted by a downstream reader model (e.g., Fusion-in-Decoder \cite{ye-etal-2023-fid}). Additionally, dense embeddings can be used for clustering, recommendation, and semantic search in non-textual domains such as code, audio, or scientific tables.

Despite these advantages, dense retrieval also poses new challenges. Indexing becomes more complex because documents must be encoded into high-dimensional vectors and stored in memory-efficient structures. Model performance is sensitive to training data quality, and domain generalization can be limited if dense representations overfit to training distributions. Moreover, most existing works on dense IR adopt encoder-only architectures, failing to utilize the powerful expression ability of Large Language Models. In hybrid systems, dense retrievers are often combined with sparse retrievers to get the best of both worlds — sparse models provide high precision for keyword-centric queries, while dense models boost recall for semantically nuanced cases.
\chapter{Background of Large Language Models(LLMs)}
\section{Overview}

Large Language Models (LLMs) are a class of deep learning models designed to process and generate human language. Trained on large-scale text corpora, LLMs learn statistical patterns of language and capture rich contextual dependencies between words, phrases, and sentences. They are typically built upon the Transformer architecture, which uses self-attention mechanisms to model relationships between all input tokens simultaneously, enabling efficient handling of long-range dependencies. During training, LLMs are optimized to predict missing or next tokens given surrounding context, resulting in representations that encode both syntactic structure and semantic meaning. The scale of training data and model parameters allows LLMs to generalize across a wide variety of linguistic tasks, making them a foundational technology in modern natural language processing research. LLMs have significantly changed the research paradigm in the AI community, and posed a profound influence on the industry.

\section{Basics of LLMs}
\subsection{Neural Language Modeling}
Neural language modeling is the task of using neural networks to estimate the probability distribution over sequences of tokens in natural language. Formally, given a sequence of tokens  $w_1, w_2, \dots, w_T$ , a language model defines a probability distribution:

\begin{equation}
    P(w_1, w_2, \dots, w_T) = \prod_{t=1}^{T} P(w_t \mid w_1, w_2, \dots, w_{t-1})
\end{equation}

where the probability of each token  $w_t$ is conditioned on all preceding tokens in the sequence. Early neural language models, such as feedforward networks, approximated  $P(w_t \mid w_{t-n}, \dots, w_{t-1})$ using a fixed-size window of previous tokens. Recurrent Neural Networks (RNNs) and Long Short-Term Memory networks (LSTMs) extended this approach by using hidden states to encode an unbounded history of past context, albeit with difficulty in capturing long-range dependencies. The introduction of the Transformer architecture further advanced neural language modeling by leveraging self-attention mechanisms to model all pairwise interactions between tokens, allowing efficient learning of both local and global contextual relationships. Training a neural language model typically involves minimizing the negative log-likelihood of observed sequences, thereby encouraging the model to assign high probabilities to naturally occurring texts:

\begin{equation}
    \mathcal{L} = - \sum_{t=1}^{T} \log P(w_t \mid w_1, w_2, \dots, w_{t-1})
\end{equation}

\subsection{Transformers and Attention Mechanisms}
\begin{figure}[htbp]
  \centering
  \includegraphics[width=0.8\linewidth]{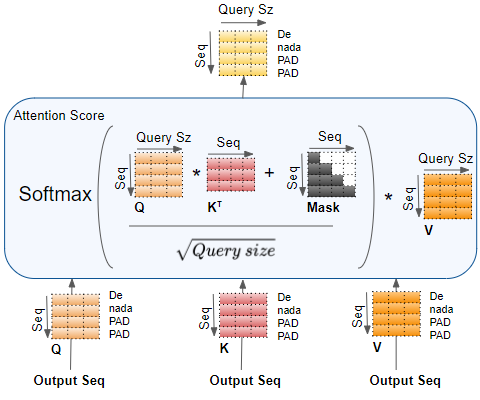}
  \caption{Illustration of the multi-head self-attention mechanism with causal masking. Graph is from the internet and available at: https://towardsdatascience.com/transformers-explained-visually-part-3-multi-head-attention-deep-dive-1c1ff1024853/. }
  \label{fig:self_attention}
\end{figure}
The Transformer architecture, introduced by Vaswani et al. \cite{vaswani2023attentionneed}, is built entirely upon attention mechanisms, dispensing with recurrence for sequence modeling. The core operation is the scaled dot-product attention, which, given a set of queries  $Q \in \mathbb{R}^{n \times d_q}$ , keys  $K \in \mathbb{R}^{n \times d_k}$ , and values  $V \in \mathbb{R}^{n \times d_v}$, computes attention outputs as:

\begin{equation}
    \text{Attention}(Q, K, V) = \text{softmax}\left( \frac{QK^\top}{\sqrt{d_k}} \right) V
\end{equation}

\begin{equation}
   Q = XW^Q, \quad K = XW^K, \quad V = XW^V 
\end{equation}

where $d_k, d_q, d_v$ are the hidden size of keys, queries and values respectively, and n is the sequence length. The softmax function ensures that the attention weights across keys for each query sum to one, enabling the model to attend differentially to different positions in the input sequence. The scaling factor  $\sqrt{d_k}$  prevents the dot products from becoming too large when  $d_k$  is high, stabilizing gradients during training.

To enhance model expressiveness, multi-head attention is employed, where the attention function is applied $h$ times in parallel with different learned linear projections, and the outputs are concatenated:

\begin{equation}
    \text{MultiHead}(X) = \text{Concat}(\text{head}_1, \dots, \text{head}_h)W^O
\end{equation}

\begin{equation}
    \text{head}_i = \text{Attention}(XW_i^Q, XW_i^K, XW_i^V) \quad\text{for i =1,2,..., h}
\end{equation}

where  $W_i^Q, W_i^K, W_i^V, W^O$ are learned parameter matrices and $X$ is the hidden states input to the attention layer.

Most modern LLMs adopt decoder-only architecture. To make it compatible with language modeling objective, a mask matrix is often added to the attention logits to selectively control which tokens each position is allowed to attend to. Before applying the softmax operation in the scaled dot-product attention, a mask  $M \in \mathbb{R}^{n \times n}$  is added to the raw attention scores and the attention becomes:

\begin{equation}
    \text{Attention}(Q, K, V) = \text{softmax} \left( \frac{QK^\top}{\sqrt{d_k}} + M \right) V
\end{equation}

 Here a causal mask is applied to prevent each position from attending to future tokens. Formally, for all positions  $i < j$ , we set  $M_{ij} = -\infty$ , and  $M_{ij} = 0$  otherwise. This ensures that the model only uses information from tokens up to the current position during prediction:

\begin{equation}
    M_{ij} =
\begin{cases}
0 & \text{if } j \leq i \\
-\infty & \text{if } j > i
\end{cases}
\end{equation}

\subsection{Modern LLMs}
Modern Large Language Models (LLMs) refer to transformer-based, decoder-only architectures trained at scale on diverse and massive textual corpora, enabling them to perform a wide range of language understanding and generation tasks with minimal task-specific supervision. A prototypical example is the GPT (Generative Pre-trained Transformer) family \cite{openai2024gpt4technicalreport}, which adopts an autoregressive causal language modeling objective and scales to hundreds of billions of parameters, demonstrating strong in-context learning capabilities. In parallel, the LLaMA series \cite{grattafiori2024llama3herdmodels} offers open-access, compute-efficient models trained on curated multilingual datasets, showing that high performance can be achieved even with fewer parameters when data quality and architectural design are optimized. An increasingly prominent trend in LLM design is the use of Mixture-of-Experts (MoE) architectures, exemplified by models like DeepSeek-V2 \cite{deepseekai2024deepseekv2strongeconomicalefficient}, which dynamically activate only a subset of expert subnetworks during inference. This approach significantly improves parameter efficiency—enabling trillions of total parameters while keeping active compute per token manageable. These modern LLMs are often further adapted using techniques like instruction tuning, reinforcement learning with human feedback (RLHF), and retrieval augmentation. Together, they represent a new era in LLM development, emphasizing not only scale, but also specialization, efficiency, and alignment with human intent.

\section{Emergent Abilities of LLMs}
Emergent abilities are those capabilities not present in smaller language models but present in larger language models \cite{wei2022emergentabilitieslargelanguage}. The existence of emergent abilities in LLMs gives us strong motivations to adopt them in downstream tasks including IR, although they seem not directly relevant.
\subsection{Few-Shot and Zero-Shot Learning}
One of the most widely studied emergent abilities of large language models is their capacity for few-shot and zero-shot learning—the ability to perform novel tasks with little or no task-specific supervision. In few-shot learning, the model is presented with a handful of input-output examples in the prompt (e.g., translations or question-answer pairs), followed by a new input for which it must generate an appropriate output. In zero-shot learning, even these demonstrations are absent; instead, the model is prompted solely with a task description or question. These behaviors are considered emergent because they typically do not appear in smaller models and only become reliable when the model reaches a sufficient scale in terms of parameters and pretraining data. For example, GPT-3 \cite{brown2020languagemodelsfewshotlearners}, with 175 billion parameters, was the first model to demonstrate strong few-shot and zero-shot performance across a wide variety of NLP benchmarks without any gradient updates, relying entirely on its internal representations and in-context pattern recognition. This capability suggests that sufficiently large language models implicitly acquire a form of meta-learning—learning how to learn—through language modeling alone.
\subsection{Chain-of-Thought Reasoning}
Chain-of-thought (CoT) reasoning refers to a language model’s ability to generate intermediate reasoning steps when solving complex tasks that involve logic, arithmetic, or multi-step inference \cite{wei2023chainofthoughtpromptingelicitsreasoning}. Rather than producing a direct answer in a single step, a model exhibiting CoT reasoning outputs a structured sequence of thoughts that mimics human-like problem solving—for example, first restating the problem, identifying relevant quantities, performing calculations, and then drawing a conclusion. This behavior is not explicitly taught during pretraining but can be elicited using carefully designed prompts, such as prefixing a question with “Let’s think step by step.” Notably, chain-of-thought reasoning is an emergent capability: it does not consistently appear in small or medium-scale models, but begins to manifest reliably only in large models (e.g., over 100B parameters). Empirical studies have shown that chain-of-thought prompting significantly improves performance on reasoning-intensive benchmarks such as GSM8K, MultiArith, and CommonsenseQA. The emergence of this ability suggests that large models can internalize abstract problem-solving procedures purely from next-token prediction training over diverse and complex textual data.
\subsection{Analogical Reasoning}
Analogical reasoning is the cognitive process of identifying structural similarities between different domains or concepts—commonly expressed in linguistic form as “A is to B as C is to ?”. In the context of large language models, analogical reasoning emerges as the model’s ability to generalize relational patterns learned from textual data and apply them to new, unseen examples. For instance, given the analogy “Paris is to France as Tokyo is to ?”, an LLM may correctly output “Japan” by recognizing the shared country–capital relationship. While small models struggle to identify such abstract relationships, larger models—especially those trained at scale—demonstrate increasing competence at both surface-level and abstract analogies. LLMs can also handle more complicated analogical reasoning, such as math word problems with similar computational graphs \cite{yang2024learninganalogyenhancingfewshot}. This capability is considered emergent because it arises spontaneously at sufficient scale, without direct supervision on analogical tasks. Empirical evaluations, such as those on the SAT analogies or relational reasoning benchmarks, show that model performance on analogical reasoning tasks improves sharply only beyond a certain parameter threshold, highlighting the non-linear nature of this emergent phenomenon.

\section{LLMs in the Context of IR}
LLMs have introduced new paradigms to Information Retrieval (IR) by unifying language understanding, reasoning, and generation within a single autoregressive framework. They can be leveraged as generative readers, especially in the case of a Retrieval-Augmented Generation (RAG) system. They can also be repurposed as retrievers based on their internal representations, usually by enabling bidirectional attention followed by further training \cite{behnamghader2024llm2veclargelanguagemodels}. Finally, an LLM can serve as both the retriever and the reader after generative representational instruction tuning \cite{muennighoff2025generativerepresentationalinstructiontuning}. As these models scale, their ability to implicitly reason about document relevance and user intent opens new directions for integrating generation and retrieval in a tightly coupled manner.

\begin{figure}[htbp]
  \centering
  \includegraphics[width=0.8\linewidth]{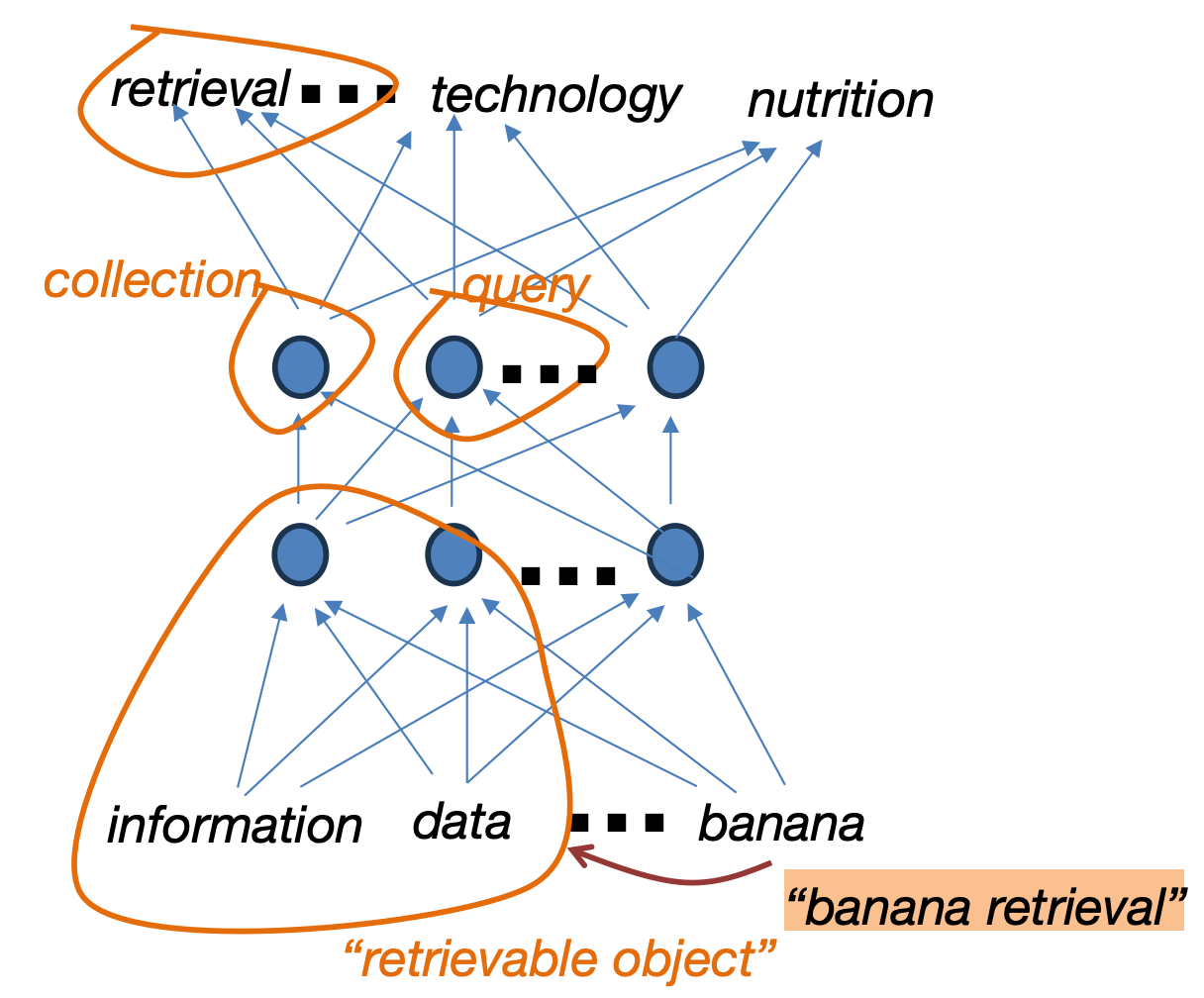}
  \caption{Illustration of the internal working mechanism of LLM as an IR task. Graph is from the slides attached to this paper \cite{IR_nodes}. }
  \label{fig:node}
\end{figure}

On the other hand, the internal working mechanism of LLMs is also related to IR. Figure \ref{fig:node} recasts the self-attention graph of a decoder-only Transformer as a miniature information-retrieval system: each token’s hidden state (blue nodes) acts simultaneously as a document and an index entry, while the directed attention weights (blue arrows) indicate retrieval scores from “query” tokens to “collection” tokens; topical cues such as retrieval, technology, and nutrition at the top of the diagram represent semantic index terms that broadcast information downstream, and the highlighted path from the composite prompt “banana retrieval” to the token banana inside the “retrievable object” contour exemplifies how relevance is established by scaled dot-product similarity. By overlaying classical IR roles — query, collection, retrievable objects, and relevance links — onto the attention mechanism, we can see that self-attention already implements a differentiable retrieval operation, so converting a large language model into an explicit dense retriever requires only minimal additional machinery.

\clearpage  
\vspace*{0pt} 

\section{The Lack of Model Interpretability Hinders the Use of LLMs on IR}

\begin{figure}[htbp]
  \centering
  \includegraphics[width=0.99\linewidth]{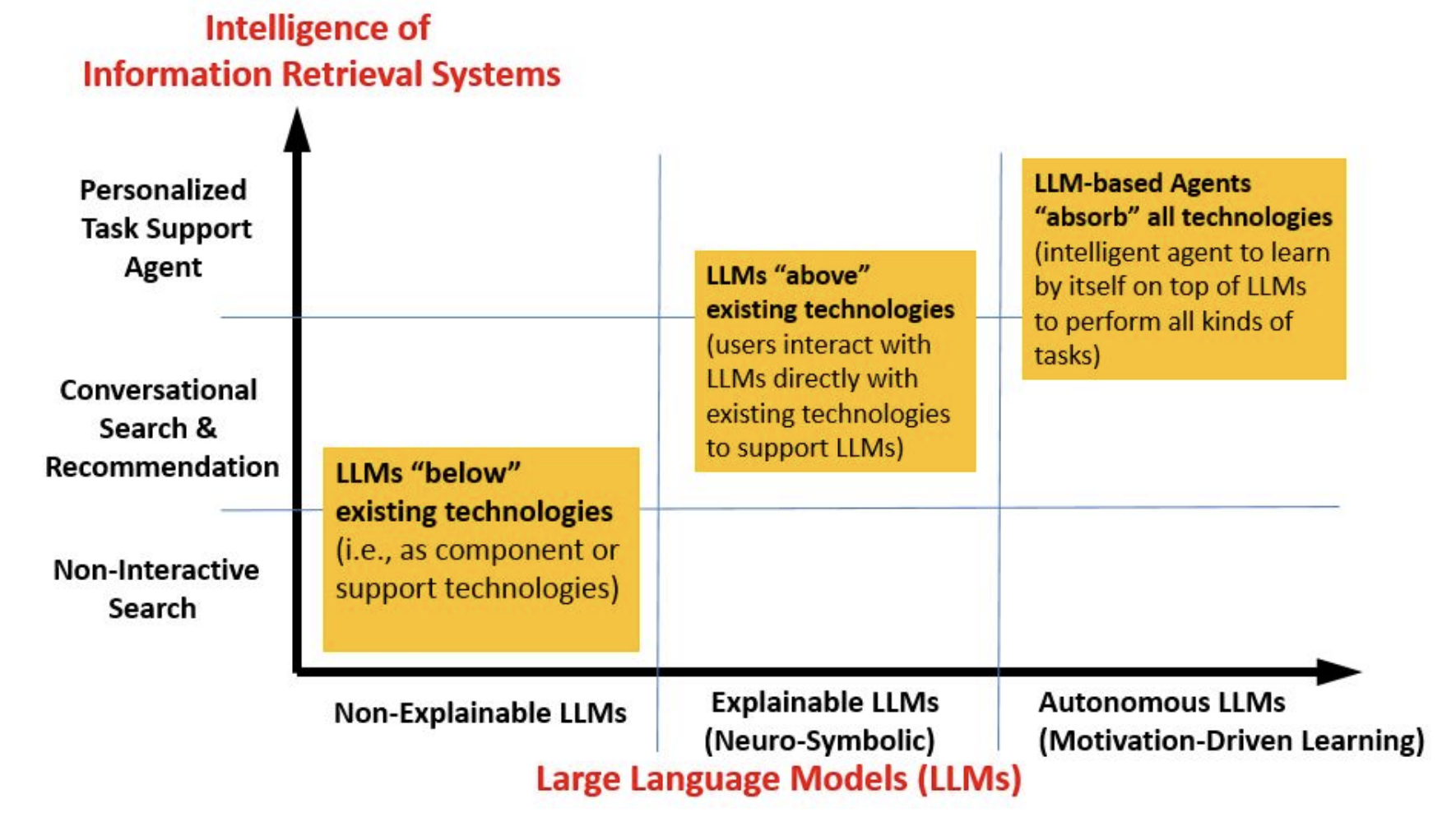}
  \caption{The future of LLMs and IR. Graph is from this paper \cite{IR_nodes}. }
  \label{fig:future}
\end{figure}

Modern LLMs are regarded lack of interpretability. In particular, the neurons in LLMs are found to be ``monosemantic" \cite{templeton2024scaling}, making it hard to interpret LLMs as in Figure \ref{fig:node}. Because ranking decisions must often be inspected by engineers, auditors, or even end-users (“Why did this document outrank that one?”), the black-box nature of LLMs stands in sharp contrast to decades of deterministic scoring functions such as BM25 or learning-to-rank with human-engineered features. Consequently, organizations with high-stakes retrieval tasks often deploy LLM components only behind retrieval-augmented guard rails, or revert to hybrid architectures where a transparent sparse retriever surfaces candidate passages that can later be sanity-checked.  Bridging this interpretability gap—through circuit-level analysis, faithful explanation generators, or intrinsically interpretable objective functions—is therefore pivotal for transferring LLM breakthroughs from research prototypes into trustworthy, production-ready IR systems, as stated in Figure \ref{fig:future}. 

\clearpage  
\vspace*{0pt}  

\chapter{Contrastive Learning Converts Large Language Models into Information Retrievers}
In this chapter, we will introduce our methodology of converting LLMs into information retrieval models through simple unsupervised contrastive learning, as well as the experimental results.

\section{Method}

\begin{figure}[htbp]
  \centering
  \includegraphics[width=0.98\linewidth]{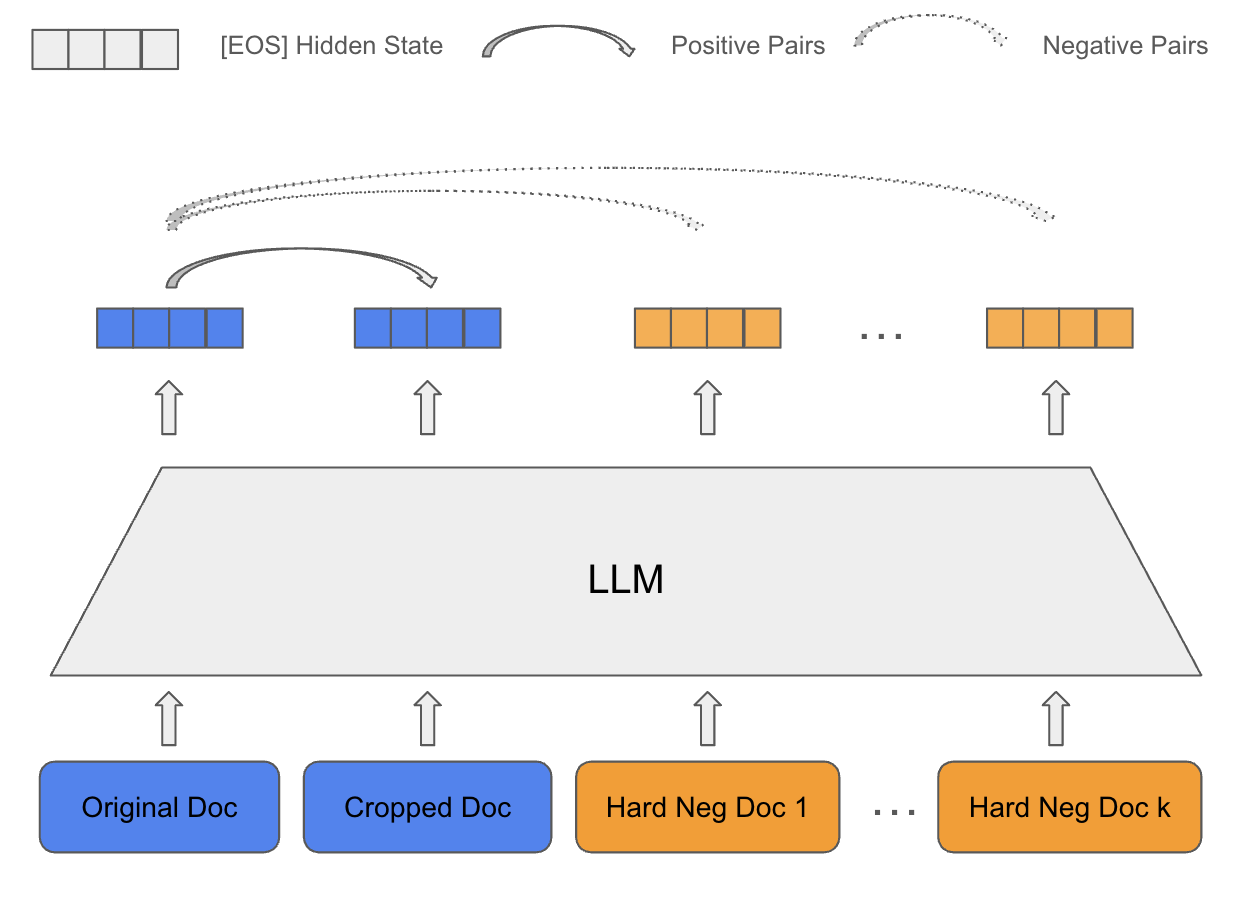}
  \caption{Contrastive training framework of our method. We use random cropping to get cropped document as positive, and use BM25 to search in the corpus to find similar pairs as hard negatives. Note that for the convenience of illustration, in-batch negatives are not drawn out in this graph.}
  \label{fig:main_graph}
\end{figure}

Given a decoder-only LLM, our goal is to convert it to an embedding model which can be used for information retrieval. To this end, we propose a simple unsupervised contrastive learning method, called LLM2IR, where we use different cropped segments of the same document as positives, and BM25-mined similar documents along with in-batch samples as negatives. Formally, for each document $d_i$ in the training corpus, we search for the most similar $K$ documents within the corpus as negatives $\{d_{i,k}^-\}_{k=1}^K$ with BM25 algorithm. Then given a batch of documents $\{{d_i}\}_{i=1}^N$, the loss function $l_i$ for the $i$-th document would be

\begin{equation}
      l_i = -\log \frac{e^{\text{sim}({d}_i, {d}_i^\prime) / \tau}}{\sum_{j=1}^N \left( e^{\text{sim}({d}_i, {d}_j^\prime) / \tau} + \sum_{k=1}^K e^{\text{sim}({d}_i, {d}_{j,k}^-) / \tau} \right)}  
\end{equation}

Where ${d}_i^\prime$ is the different cropped segment of ${d}_i$ and $\tau$ is the temperature. $Sim(\cdot, \cdot)$ is defined as the cosine similarity of two document embeddings, where we use the last hidden state of the \texttt{[EOS]}\footnote{For phi-3 models, we use \texttt{$<$/s$>$} instead of \texttt{$<$|endoftext|$>$} as the \texttt{[EOS]} token.} token in our decoder-only models. 

\section{Training Data Construction}
A key step in unsupervised contrastive learning is to construct training data pairs from a single document. Earlier methods include random word deletion, replacement or masking as data augmentation. Inverse Cloze Task (ICT) \cite{lee-etal-2019-latent} creates two distinct views of a document where the first view is generated by randomly selecting a span of tokens from a text segment, with the second view consisting of the remaining tokens from that span. Title Language Models \cite{titleIR} used the title of the passage as the query. SimCSE \cite{gao2022simcsesimplecontrastivelearning} adopts two independent forward pass with different dropout masks to introduce perturbation. Empirically, we find that the latter gives a poor performance with mistral v0.1 backbone, and simple random cropping gives the best performance, as shown in Table \ref{table:simCSE}.

\begin{table}[htbp]
\centering
\caption{Performance comparison of different data augmentation techniques on a subset of BEIR benchmark.}
\label{table:simCSE}
\renewcommand{\arraystretch}{1.2} 
\vspace{1em}
\begin{tabular}{lcc}
\hline
& \textbf{SimCSE dropout masks} & \textbf{Random cropping } \\
\hline
SciFact & 17.0 & \textbf{72.2} \\
ArguAna & 11.9 & \textbf{51.6} \\
NFCorpus & 2.6 & \textbf{36.3} \\
QuoraRetrieval & 52.2 & \textbf{85.3} \\
SCIDOCS & 1.2 & \textbf{20.7} \\
FiQA2018 & 2.0 & \textbf{40.8} \\
TRECCOVID & 15.4 & \textbf{53.2} \\
\hline
\end{tabular}
\end{table}

\section{Experiments}
In this section, we present the main experiment results of \textsc{LLM2IR} on LoCo, LongEmbed and BEIR benchmarks.

\subsection{Setups}
\textbf{Training Procedure.} We choose Mistral v0.2 and Mistral v0.1 models \cite{jiang2023mistral7b} to be converted. We train each of our models on Wikitext-103 dataset \cite{wikitext103} for 1 epoch, following the setting in LLM2Vec. This training process comprises 460 steps with a global size of 64 on 8 A6000 GPUs using Deepspeed ZeRO stage 3. Our models are trained with LoRA \cite{hu2021loralowrankadaptationlarge} using AdamW optimizer \cite{AdamW} and a learning rate of 1e-4. We set $K$ = 7 for top-k negative mining for each passage. Each anchor document $d_i$ is a randomly-cropped consecutive segment of 64 tokens from the original passage, and positive and negative documents are truncated to 512 tokens. We add a prefix ``\textit{Query: }" to each anchor and ``\textit{Passage: }" to the positive and negative documents. Temperate $\tau$ is set to 0.05. Brain floating point (bfloat16) quantization, FlashAttention-2 \cite{dao2023flashattention2fasterattentionbetter} and gradient checkpointing are applied. 

\textbf{Baseline methods.} We take LLM2Vec \cite{LLM2Vec} and E5-Mistral \cite{E5-mistral} as our baseline methods. LLM2Vec finetunes model based on Mistral v0.2, and E5-Mistral does it on Mistral v0.1. We take the unsupervised results from LLM2Vec for a fair comparison. Note that E5-Mistral trains model with labeled data. Even so, we'll show that our method raises comparable results to theirs.

\subsection{Long Context Retrieval: LoCo and LongEmbed}
\textbf{Dataset Description.} LoCo benchmark \cite{Loco} is a collection of a 12 tasks constructed to measure long-context retrieval where chunking is not possible or not effective. LoCo incorporates data from various established long-context benchmarks, such as Tau Scrolls \cite{tauscrolls}, LongBench \cite{bai2024longbenchbilingualmultitaskbenchmark}, and QASPER \cite{Qasper}. Additionally, it includes several domain-specific datasets not initially designed for retrieval, including CourtListener, the Australian Legal Court Reports dataset, and the StackOverflow forum. LongEmbed benchmark \cite{zhu2024longembedextendingembeddingmodels} consists of two synthetic tasks and four meticulously selected real-world tasks, each featuring documents of different lengths and scattered target information. In this work, we only consider real-world tasks.

\clearpage  
\vspace*{0pt} 

\begin{table}[htbp]
\centering
\caption{LoCo benchmark statistics.}
\label{table:loco_stats_test_only}
\renewcommand{\arraystretch}{1.2}
\resizebox{\textwidth}{!}{%
\begin{tabular}{@{}l r r r r@{}}
\toprule
\textbf{Dataset}                    & \textbf{\# Queries} & \textbf{\# Docs} & \textbf{Avg.\ Query Words} & \textbf{Avg.\ Doc Words} \\
\midrule
SummScreenFD                       & 338    & 338    & 590   & 30,792 \\
Gov.\ Report                       & 972    & 972    & 3,871 & 55,280 \\
QMSum                              & 272    & 272    & 430   & 58,129 \\
QASPER Title                       & 416    & 416    & 71    & 22,315 \\
QASPER Abstract                    & 416    & 416    & 931   & 22,315 \\
MultiFieldQA                       & 30     & 30     & 62    & 29,465 \\
2WikimQA                           & 60     & 60     & 69    & 37,867 \\
Passage Retrieval                  & 60     & 60     & 840   & 35,814 \\
CourtListener (Plain Text)         & 2,000  & 2,000  & 146   & 48,190 \\
CourtListener (HTML)               & 2,000  & 2,000  & 146   & 57,028 \\
Australian Legal Case Report       & 770    & 770    & 14,986 & 47,536 \\
StackOverflow                      & 400    & 7,741  & 758   & 4,544  \\
\bottomrule
\end{tabular}%
}
\end{table}

\begin{table}[htbp]
\centering
\caption{Statistics of the LongEmbed real‐world datasets.}
\label{table:real_tasks_stats}
\renewcommand{\arraystretch}{1.2}
\resizebox{\textwidth}{!}{%
\begin{tabular}{@{}l r r r r@{}}
\toprule
\textbf{Dataset}       & \textbf{\# Queries} & \textbf{\# Docs} & \textbf{Avg.\ Query Words} & \textbf{Avg.\ Doc Words} \\
\midrule
NarrativeQA       & 10,449 & 355 & 9   & 50,474 \\
QMSum             & 1,527  & 197 & 71  & 10,058 \\
2WikiMultihopQA   & 300    & 300 & 12  & 6,132  \\
SummScreenFD      & 336    & 336 & 102 & 5,582  \\
\bottomrule
\end{tabular}%
}
\end{table}

\textbf{Experiment Results.} Table \ref{table:embeddings_tasks} shows results on the real-world tasks from LongEmbed benchmark. On QMSum, SummScreenFD and NarrativeQA tasks, LLM2IR with Mistral v0.2 significantly outperforms the baseline methods, and LLM2IR with Mistral v0.1 raises a performance even better than LLM2Vec with Mistral v0.2. Table \ref{table:embeddings_tasks_new} shows results on 12 tasks on LoCo Benchmark. Similarly, LLM2IR with Mistral v0.2 outperforms the baselines. Even LLM2IR with Mistral v0.1 results a comparable performance to E5-Mistral, which is significantly better than LLM2Vec.

\begin{table}[htbp]
\centering
\caption{Experiment results on LongEmbed Benchmark.}
\label{table:embeddings_tasks}
\renewcommand{\arraystretch}{1.2} 
\begin{tabular}{lcccc}
\hline
& \textbf{\model (v0.2)} & \textbf{\model (v0.1)} & \textbf{LLM2Vec} & \textbf{E5-Mistral} \\
\hline
QMSum & \textbf{49.7} & 42.2 & 38.6 & 43.6 \\
2WikimQA & 70.5 & 53.9 & 40.2 & \textbf{82.0} \\
SummScreenFD & \textbf{98.3} & 94.8 & 89.2 & 96.8 \\
NarrativeQA & \textbf{59.8} & 51.2 & 38.1 & 44.6 \\
\hline
Avg. & \textbf{69.6} & 60.5 & 51.5 & 66.8 \\
\hline
\end{tabular}
\end{table}

\begin{table}[htbp]
\centering
\caption{Experiment results on LoCo Benchmark.}
\label{table:embeddings_tasks_new}
\renewcommand{\arraystretch}{1.2} 
\begin{tabular}{lcccc}
\hline
& \textbf{LLM2IR(v0.2)} & \textbf{LLM2IR(v0.1)} & \textbf{LLM2Vec} & \textbf{E5-Mistral} \\
\hline
SummScreenFD & \textbf{98.2} & 94.7 & 91.4 & 95.9 \\
Gov. Report & \textbf{98.8} & 97.1 & 92.1 & 98.3 \\
QMSum & \textbf{75.0} & 67.1 & 64.8 & 46.8 \\
QASPER title & 93.0 & 86.9 & 63.8 & \textbf{98.4} \\
QASPER abstract & \textbf{99.8} & 98.3 & 83.6 & \textbf{99.8} \\
MultifieldQA & 92.4 & 92.4 & 84.7 & \textbf{93.5} \\
2WikimQA & 78.6 & 58.3 & 62.0 & \textbf{88.3} \\
Passage Retrieval & \textbf{44.0} & 22.5 & 36.3 & 35.3 \\
Legal Case Reports & \textbf{58.8} & 54.0 & 28.6 & 49.5 \\
C.L. (HTML) & \textbf{34.3} & 33.5 & 16.3 & 33.9 \\
C.L. (Plain Text) & \textbf{34.6} & 34.0 & 21.6 & \textbf{34.6} \\
Stackoverflow & 81.5 & 80.9 & 64.9 & \textbf{82.7} \\
\hline
Avg. & \textbf{74.1} & 68.3 & 59.2 & 71.4 \\
\hline
\end{tabular}
\end{table}

\subsection{Short Context Retrieval: BEIR} 
\textbf{Dataset Description.} Benchmarking-IR (BEIR) 
 \cite{thakur2021beirheterogenousbenchmarkzeroshot} is a robust and heterogeneous evaluation benchmark for information retrieval consisting of publicly available datasets from diverse text retrieval tasks and domains. It aims to provide a unified framework for evaluating the generalization capabilities of IR models, allowing the test of zero-shot capabilities of IR models.

\begin{table}[htbp]
\centering
\caption{Statistics of BEIR benchmark.}
\label{table:test_stats_overview}
\renewcommand{\arraystretch}{1.2}
\resizebox{\textwidth}{!}{%
\begin{tabular}{@{}l r r r r@{}}
\toprule
\textbf{Dataset}    & \textbf{\# Test Queries} & \textbf{\# Test Docs} & \textbf{Avg.\ Query Len.} & \textbf{Avg.\ Doc Len.} \\
\midrule
MS MARCO            & 6,980    & 8,841,823   & 5.96  & 55.98  \\
TREC-COVID          & 50       & 171,332     & 10.60 & 160.77 \\
NFCorpus            & 323      & 3,633       & 3.30  & 232.26 \\
BioASQ              & 500      & 14,914,602  & 8.05  & 202.61 \\
NQ                  & 3,452    & 2,681,468   & 9.16  & 78.88  \\
HotpotQA            & 7,405    & 5,233,329   & 17.61 & 46.30  \\
FiQA-2018           & 648      & 57,638      & 10.77 & 132.32 \\
Signal-1M (RT)      & 97       & 2,866,316   & 9.30  & 13.93  \\
TREC-NEWS           & 57       & 594,977     & 11.14 & 634.79 \\
Robust04            & 249      & 528,155     & 15.27 & 466.40 \\
ArguAna             & 1,406    & 8,674       & 192.98& 166.80 \\
Touché-2020         & 49       & 382,545     & 6.55  & 292.37 \\
CQADupStack         & 13,145   & 457,199     & 8.59  & 129.09 \\
Quora               & 10,000   & 522,931     & 9.53  & 11.44  \\
DBPedia             & 400      & 4,635,922   & 5.39  & 49.68  \\
SCIDOCS             & 1,000    & 25,657      & 9.38  & 176.19 \\
FEVER               & 6,666    & 5,416,568   & 8.13  & 84.76  \\
Climate-FEVER       & 1,535    & 5,416,593   & 20.13 & 84.76  \\
SciFact             & 300      & 5,183       & 12.37 & 213.63 \\
\bottomrule
\end{tabular}%
}
\end{table}

\textbf{Experiment Results.} As shown in Table \ref{table:embeddings_tasks_beir}, LLM2IR outperforms LLM2Vec and BM25 baselines. This suggests that LLM2IR is a competitive and robust retrieval method across diverse retrieval tasks, especially in challenging settings.

\begin{table}[htbp]
\centering
\caption{Experiment results on BEIR benchmark.}
\label{table:embeddings_tasks_beir}
\renewcommand{\arraystretch}{1.2} 
\begin{tabular}{lccc}
\hline
& \textbf{LLM2IR (v0.2)} & \textbf{BM25} & \textbf{LLM2Vec} \\
\hline
SciFact & \textbf{72.5} & 66.5 & 68.67 \\
ArguAna & 49.5 & 31.5 & \textbf{57.48} \\
NFCorpus & \textbf{35.2} & 32.5 & 27.16 \\
QuoraRetrieval & \textbf{85.5} & 78.9 & 84.4 \\
SCIDOCS & \textbf{21.5} & 15.8 & 15.35 \\
FiQA2018 & \textbf{44.1} & 23.6 & 27.24 \\
TRECCOVID & 52.3 & \textbf{65.6} & 55.66 \\
NQ & 29.9 & 32.9 & \textbf{34.16} \\
FEVER & 58.5 & \textbf{75.3} & 45.11 \\
Touche2020 & 15.1 & \textbf{36.7} & 6.54 \\
ClimateFEVER & 21.8 & 21.3 & \textbf{22.97} \\
DBPedia & \textbf{42.7} & 31.3 & 25.48 \\
HotpotQA & 54.1 & \textbf{60.3} & 54.54 \\
\hline
BEIRAvg & \textbf{44.8} & 44.0 & 40.4 \\
\hline
\end{tabular}
\end{table}

\section{Discussion}
\subsection{Carefully-Calibrated Finetuning Pipeline Is Not Necessary}
Previous works on converting decoder-only LLMs into information retrievers (or general embedding models) usually adopt complicated techniques. For example, LLM2Vec uses a three-stage training pipeline including enabling bidirectional attention, masked next token prediction training and contrastive learning; GRIT uses a unified representational instruction tuning and generative instruction tuning with human-constructed instruction data. Our approach empirically proves that these carefully designed training techniques are not necessary, and simple unsupervised contrastive learning on self-cropped data is good enough. Notably, we find that enabling bidirectional attention, which is declared an important operation in both LLM2Vec and GRIT, does not significantly affect the performance of converted models on IR tasks, probably because that the \texttt{[EOS]} token itself can attend to all tokens and therefore gather information from them. In summary, our findings challenge the prevailing assumption that complex architectural modifications and supervised instruction tuning are essential, showing instead that a minimalist, unsupervised approach can achieve competitive retrieval performance.

\subsection{Hard Negative Mining is Crucial}
We empirically observed that the mining of hard negative samples significantly influences the IR performance of converted models. Without hard negatives, the training loss decreases to 0 after the first batch, suggesting that no additional information is learned from the training data. This is consistent with previous contrastive learning works, and in our case it may be due to the use of random cropping, where the model can identify positive pairs by recognizing overlapped segments. While with hard negative samples, which is mined by BM25 which relies on key word matching, the model have to tell apart those hard negatives from positives, so that models would learn more semantic similarities.  

\chapter{LONG CONTEXT Large Language Models ARE BETTER INFORMATION RETRIEVERs}

In this chapter, we will show that when the input length touchs the ceiling of the retrieval model, the IR performance drops significantly. In other words, models with longer context length tends to raise a better IR result given other factors kepted the same.

\section{Setups}
In this chapter, we explore the effect that model context length has on IR task performance, based on the method proposed in the last chapter. To keep the model context length the only difference between model pairs, we choose Phi3-mini-128k / Phi3-mini-4k \cite{abdin2024phi3technicalreporthighly} and Mistral-v0.1-yarn-128k \cite{peng2023yarnefficientcontextwindow}/ Mistral-v0.1-8k as our base models and test the converted models on LoCo benchmark. Here, Phi3-mini-128k is finetuned from Phi3-mini-4k using LongRope \cite{ding2024longropeextendingllmcontext} on minimal data. Similarly, Mistral-v0.1-yarn-128k is finetuned from Mistral-v0.1-8k using YaRN \cite{peng2023yarnefficientcontextwindow}. Note that for fair comparison, the input texts are truncated to the shorter length of two models, i.e. the maximal input length is 4096 for both Phi3-mini-128k and Phi3-mini-4k and 8192 for both Mistral v0.1-yarn-128k and Mistral v0.1-8k, ensuring that the inputs to two models are the same. All other setups are kepted the same as the experiments in the last chapter.

\section{Experiment Results}
As shown in Table \ref{table:context_length_comparison}, we can see that in experiments with both model pairs, the model with longer context length outperforms the model with shorter context length significantly on most tasks.

\clearpage  
\vspace*{0pt}  

\begin{table}[htbp]
\centering
\caption{Performance Comparison of models with different context length on LoCo benchmark. 
Note that the input texts are truncated to the shorter length of two models, i.e., the maximal input length is 4096 for both Phi3-mini-128k and Phi3-mini-4k, and 8192 for both Mistral v0.1-yarn-128k and Mistral v0.1-8k.}
\label{table:context_length_comparison}
\renewcommand{\arraystretch}{1.2}
\vspace{1em}

\resizebox{\textwidth}{!}{%
\begin{tabular}{lcc}
\toprule
 & \textbf{LLM2IR (Phi3-mini, 128k)} & \textbf{LLM2IR (Phi3-mini, 4k)} \\
\midrule
SummScreenFD       & \textbf{94.7} & 89.7 \\
Gov. Report        & \textbf{98.3} & 96.9 \\
QMSum              & \textbf{64.3} & 60.8 \\
QASPER title       & \textbf{90.4} & 82.7 \\
QASPER abstract    & \textbf{99.8} & 96.5 \\
MultifieldQA       & \textbf{92.1} & 91.9 \\
2WikimQA           & \textbf{78.7} & 57.5 \\
Passage Retrieval  & \textbf{40.0} & 16.2 \\
Legal Case Reports & \textbf{45.6} & 37.3 \\
C.L. (HTML)        & 25.9 & \textbf{26.1} \\
C.L. (Plain Text)  & \textbf{26.6} & 26.2 \\
Stackoverflow      & 79.2 & \textbf{79.8} \\
\midrule
Avg.               & \textbf{69.6} & 63.5 \\
\bottomrule
\end{tabular}%
}
\vspace{2em}

\resizebox{\textwidth}{!}{%
\begin{tabular}{lcc}
\toprule
 & \textbf{LLM2IR (Mistral v0.1-yarn, 128k)} & \textbf{LLM2IR (Mistral v0.1, 8k)} \\
\midrule
SummScreenFD       & \textbf{96.3} & 94.7 \\
Gov. Report        & \textbf{97.9} & 97.1 \\
QMSum              & 66.6 & \textbf{67.1} \\
QASPER title       & \textbf{90.6} & 86.9 \\
QASPER abstract    & \textbf{99.4} & 98.3 \\
MultifieldQA       & \textbf{93.8} & 92.4 \\
2WikimQA           & \textbf{61.5} & 58.3 \\
Passage Retrieval  & \textbf{22.7} & 22.5 \\
Legal Case Reports & \textbf{56.5} & 54.0 \\
C.L. (HTML)        & \textbf{35.1} & 33.5 \\
C.L. (Plain Text)  & \textbf{35.6} & 34.0 \\
Stackoverflow      & \textbf{81.6} & 80.9 \\
\midrule
Avg.               & \textbf{69.8} & 68.3 \\
\bottomrule
\end{tabular}%
}

\end{table}

\section{Discussion}
\begin{figure}[htbp]
  \centering
  \includegraphics[width=0.8\linewidth]{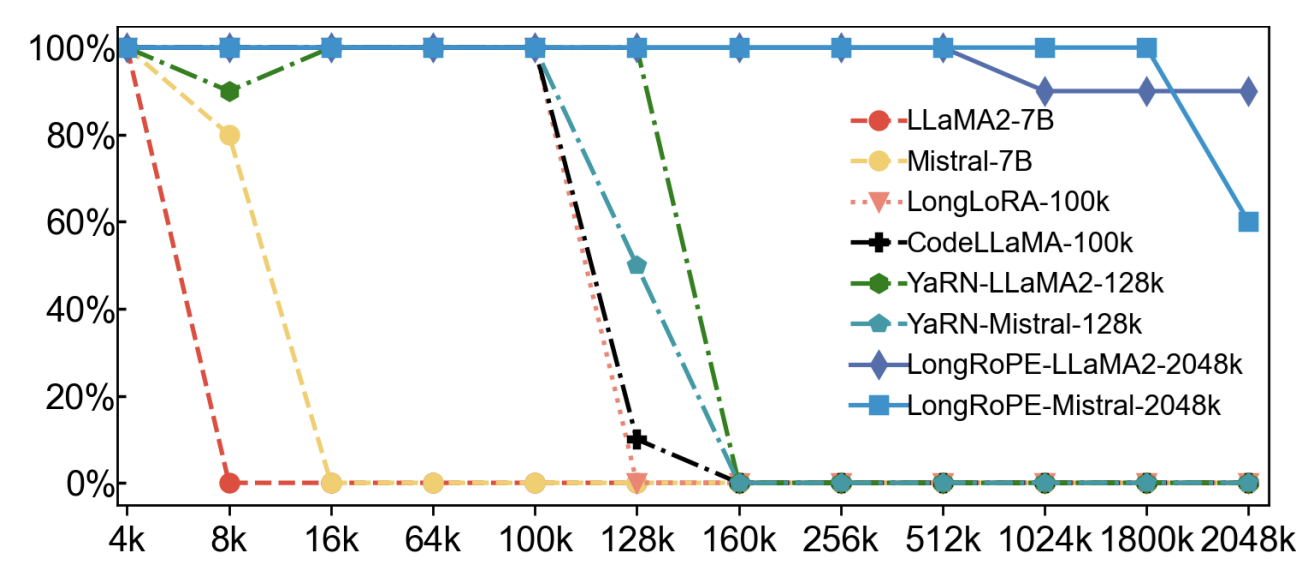}
  \caption{IR performance on Passkey dataset \cite{zhu2024longembedextendingembeddingmodels} with different models. Graph is from LongRope paper \cite{ding2024longropeextendingllmcontext}.}
  \label{fig:passkey}
\end{figure}

In our experiments, we observed that the models with shorter context length raises worse IR performance. According to Table \ref{table:loco_stats_test_only}, we can infer that most documents are truncated to their maximal length (4k/8k) since the average document lengths are significantly larger than that. Interestingly, in LongRope paper, the authors also observed similar phenomena. As shown in Figure \ref{fig:passkey}, the IR performance significantly drops when input length touchs the maximal model context length (for example, the performance of LongRoPE-Mistral-2048k drops from 100\% to 60\% when input length goes from 1800k to 2048k, and the performance of YaRN-Mistral-128k drops from 100\% to 50\% when input length goes from 100k to 128k). This performance decay indicates that it is wise to choose a retriever model whose context length is significantly longer than input length to achieve the best IR performance.

\clearpage  
\vspace*{0pt}  

\chapter{Future Works}
\section{Better Data Augmentation Method}
In this work, we used simple random cropping to create positive query-document pairs for contrastive learning. We empirically find that random dropout masks used in SimCSE raises a much worse performance than random cropping. 
However, random cropping may not be the optimal data construction method, as the cropped segments are overlapped with the original document. This could be eased by methods like Inverse Cloze Task, however, the resulted pairs are still different from real query-document pairs in semantics.

Most previous works use existing query-document pair data. For example, DPR \cite{karpukhin2020densepassageretrievalopendomain} uses user questions as queries and chunked passages as documents for contrastive learning, which is closer to the test settings. However, it relies on labeled data and therefore is not scalable for larger models. A possible future direction is to figure out how to automatically construct queries similar to user questions based on given documents with the help of lightweight Information Extraction techniques to construct large-scale training datasets.

\section{Theoretical Explanation of the Effect of Context Length}

In chapter 5, we observed an important phenomenon about the influence that model context length has on the IR performance. However, we cannot find a satisfactory theoretical explanation of it till now. 

\begin{figure}[htbp]
  \centering
  \includegraphics[width=0.8\linewidth]{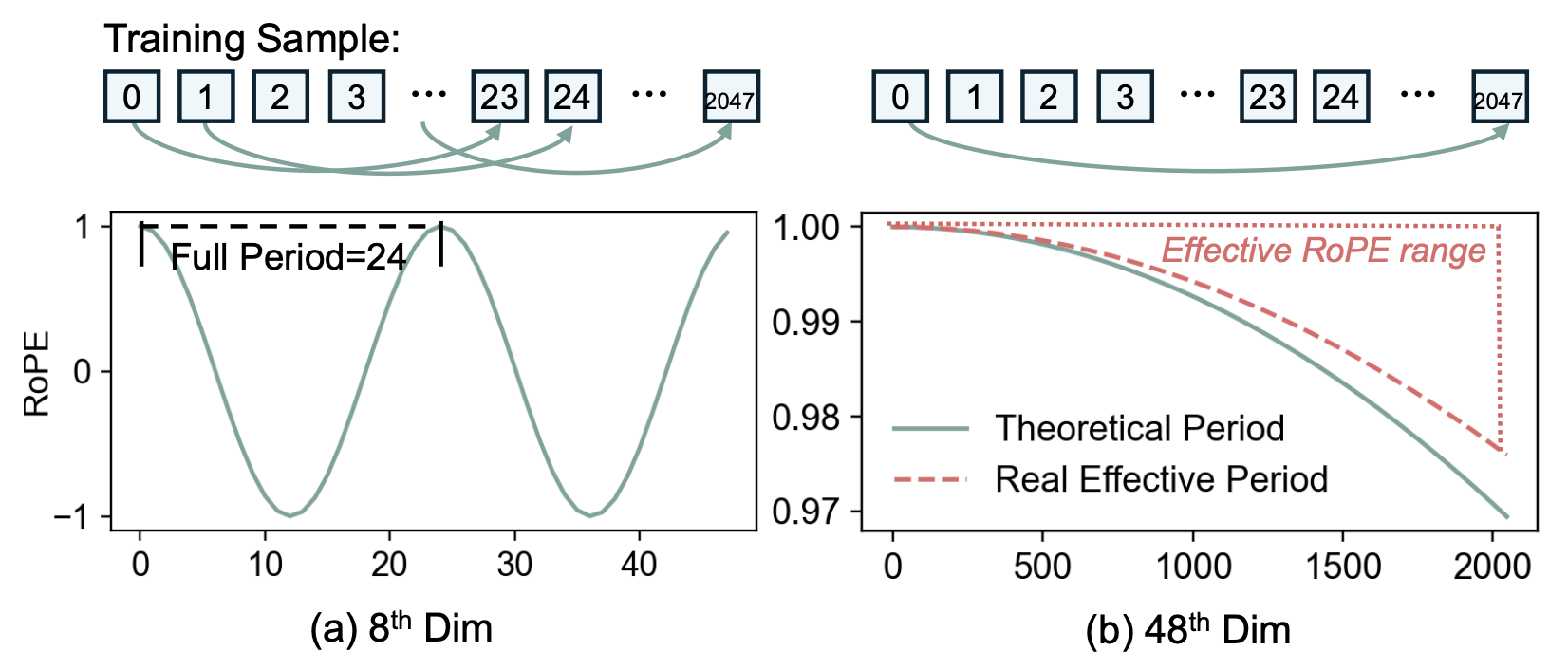}
  \caption{The higher dimension of RoPE is under-trained, resulting a shorter effective model context length. Graph is from LongRoPE2 \cite{shang2025longrope2nearlosslessllmcontext}. }
  \label{fig:longrope}
\end{figure}

A possible reason is related to under-training of high-index positions. Even after context-extension fine-tuning, tokens near the maximum length receive far fewer gradient updates. LongRoPE2 \cite{shang2025longrope2nearlosslessllmcontext} measures this and shows that higher RoPE dimensions remain under-trained, as shown in Figure \ref{fig:longrope}. This can explain why IR performance drops when input length touches the context ceiling, as it surpasses the effective context length. Another possible factor is the change of rotary angle $\theta$, as it modifies the wavelength of the input sequences. Different IR tasks may rely more on higher (local) or lower frequency (global / document-wise) information. 

\clearpage  
\vspace*{0pt}  

\chapter{Conclusion}

This thesis set out to narrow the gap between cutting-edge decoder-only large language models and production-ready dense retrievers, doing so without large-scale supervision or elaborate engineering.  By introducing LLM2IR, we showed that a single unsupervised contrastive objective — augmented only with naïve document cropping and BM25-mined hard negatives — can reliably transform diverse backbones such as Mistral and Phi-3 into competitive retrieval models.  Across long-context benchmarks (LoCo and LongEmbed) the converted models surpassed considerably more sophisticated baselines, and on the heterogeneous BEIR suite they remained on par with, and sometimes ahead of, systems trained with explicit relevance labels.  An additional empirical finding is that expanding the context window confers a tangible edge: when all other variables are held constant, 128k-token variants consistently outperform their 4k and 8k counterparts, implying that retrieval accuracy is governed not only by representational capacity but also by the distance over which attention can propagate relevance cues.

For practitioners, the most immediate takeaway is that simplicity can trump sophistication.  The proposed recipe obviates multi-stage alignment pipelines and permits reuse of the very decoder-only LLM already deployed for text generation, which in turn reduces latency, eases maintenance, and avoids costly model duplication.  Moreover, our experiments suggest that retrieval pipelines should be designed with ample head-room in context length; otherwise, once sequences reach the model’s ceiling, accuracy degrades precipitously.  Because LLM2IR operates entirely on unlabelled text, it can be scaled to specialised domains — legal, biomedical, or code — where supervised pairs are scarce or expensive to curate.  The approach therefore offers a pragmatic path toward domain-specific retrieval with minimal annotation overhead and a modest compute footprint: fine-tuning LoRA adapters for a 7-billion-parameter model completed in under eight hours on a single consumer GPU, yet delivered double-digit gains over BM25 in six BEIR tasks.

Despite these encouraging results, several caveats remain.  The reliance on random cropping, while computationally attractive, may introduce spurious positives and fails to reflect authentic information-seeking behaviour; preliminary manual inspection revealed that roughly 12\% of positive pairs share little topical overlap beyond surface-level vocabulary.  All experiments tuned only a narrow rank - 1.5\% subset of parameters via LoRA, which leaves unanswered whether full-parameter training could unlock additional performance or, conversely, accelerate over-fitting — especially on shorter contexts.  Furthermore, the “context-length cliff” documented in Section 6.2 is still theoretically opaque: we can conjecture that under-training of late positions or positional-bias misalignment are contributing factors, but a principled mechanistic account is missing.  Finally, our evaluation focused on English corpora and single-step retrieval; multilingual and multi-hop settings may surface failure modes not captured here.

Building on these findings, several research avenues appear promising.  First, synthetic-query generation — guided either by large language models or retrieval-augmented generation — could replace random cropping with semantically richer positives, potentially narrowing the domain shift between training and inference.  Second, curriculum schedules that expose the model to progressively longer spans may mitigate edge-of-window degradation and shed light on the learning dynamics behind the context-length cliff.  A third line of work is end-to-end optimization that jointly trains retrieval and generation in a single stage, unifying contrastive and language-model objectives; preliminary ablations suggest that such co-training could close the remaining gap to fully supervised retrievers while preserving the model’s generative fluency.  Beyond these immediate steps, exploring cross-modal retrieval (text-to-image, code-to-text), evaluating energy and memory efficiency on edge devices, and extending the framework to multilingual or low-resource languages would all contribute to a more comprehensive understanding of how unsupervised LLM adaptation can serve the next generation of search systems.





%
\bibliographystyle{IEEE_ECE}
\bibliography{thesisrefs}  

%
\appendix


\backmatter

\end{document}